# Electronic Lieb lattice signatures embedded in two-dimensional polymers with square lattice

Yingying Zhang,[a] Shuangjie Zhao,[a] Miroslav Položij,[abc] Thomas Heine*[abcd]

[a] Chair of Theoretical Chemistry, Technische Universität Dresden, Bergstrasse 66, 01069 Dresden, Germany

[b] Helmholtz-Zentrum Dresden-Rossendorf, HZDR, Bautzner Landstr. 400, 01328 Dresden, Germany

[c] Center for Advanced Systems Understanding, CASUS, Untermarkt 20, 02826 Görlitz, Germany

[d] Department of Chemistry and ibs for nanomedicine, Yonsei University, Seodaemun-gu, Seoul 120-749, Republic of Korean.

E-mail: thomas.heine@tu-dresden.de

**ABSTRACT**

Exotic band features, such as Dirac cones and flat bands, arise directly from the lattice symmetry of materials. The *Lieb* lattice is one of the most intriguing topologies, because it possesses both Dirac cones and flat bands which intersect at the Fermi level. However, materials with *Lieb* lattice remain experimentally unreached. Here, we explore two-dimensional polymers (2DPs) derived from zinc-phthalocyanine (ZnPc) building blocks with a square lattice (*sql*) as potential electronic *Lieb* lattice materials. By systematically varying the linker lengths (ZnPc-$x$P), we found that some ZnPc-$x$P exhibit a characteristic *Lieb* lattice band structure. Interestingly though, *fes* bands are also observed in ZnPc-$x$P. The coexistence of *fes* and *Lieb* in *sql* 2DPs challenges the conventional perception of the structure-electronic structure relation. In addition, we show that manipulation of the Fermi level, achieved by electron removal or atom substitution, effectively preserves the unique characteristics of *Lieb bands*. Chern number calculations confirm the non-trivial nature of the *Lieb* Dirac bands. Our discoveries provide a fresh perspective on 2D polymers and redefine the search for *Lieb* lattice materials into a well-defined chemical synthesis task.

**INTRODUCTION**

Exotic electronic structures, exemplified by Dirac cones and flat bands, have emerged as a focal point in contemporary research due to their unique electronic properties, including exotic charge carrier mobilities and the induction of topological effects.[1,2] An ample example of these exotic electronic structures are the Dirac cones in honeycomb (*hcb*) lattice, which were first predicted theoretically and only later gained importance with the discovery of graphene.[3,4] Since then, graphene has found many applications in electronic devices, high-speed transistors, spintronics, photonics and optoelectronics.[5] All these applications are possible thanks to the Dirac cone, characterized by crossing bands with linear dispersion intersecting at $K$ point of the Brillouin zone. This implies the existence of massless electrons from a non-relativistic perspective, consequently leading to exceedingly high electron mobility and topological effects.[4,6] On the other hand, flat (dispersionless) bands are characterized by electrons with extraordinarily large effective masses and energies that are independent of the carrier momentum.[2,7,8] Partially filled flat bands can then result in novel phases of matter, such as superconductivity, magnetism, and metal-insulator transitions.[9]

The relationship between honeycomb lattice and Dirac cones can be generalized to a statement that electronic structure features arise directly from the lattice symmetry of the materials.[10] Notably, many of the distinctive electronic features are shared between very different lattices, with Dirac cones appearing e.g. in kagome (*kgm*), *hcb*, *fes* and *Lieb* lattices, and flat bands e.g. in *kgm* and *Lieb*.[10–12] Among these, the *Lieb* lattice signature electronic structure is a very interesting one because it contains flat bands exactly crossing the Dirac cone (denoted as "*Lieb bands*" in the remaining manuscript) (Figure 1).[7] These bands may be interesting from the viewpoint of electronic topology, since the Dirac bands in an ideal *Lieb* lattice (the corner and edge sites are in the same energy, dE=0) are topologically non-trivial and the contact between them and the flat bands is protected by the real-space topology. When dE≠0, where the corner and edge sites are in different energies, the flat band becomes topologically non-trivial, while one of the Dirac bands becomes trivial with a band gap between it and the flat band being opened.[10,13,14] Theoretical predictions using the Tight-Binding (TB) model show that *Lieb bands* require ideal lattice symmetry and strict conditions on state energies.[13,15,16] Because of these very strict criteria, the electronic *Lieb* lattice has rarely been achieved experimentally.

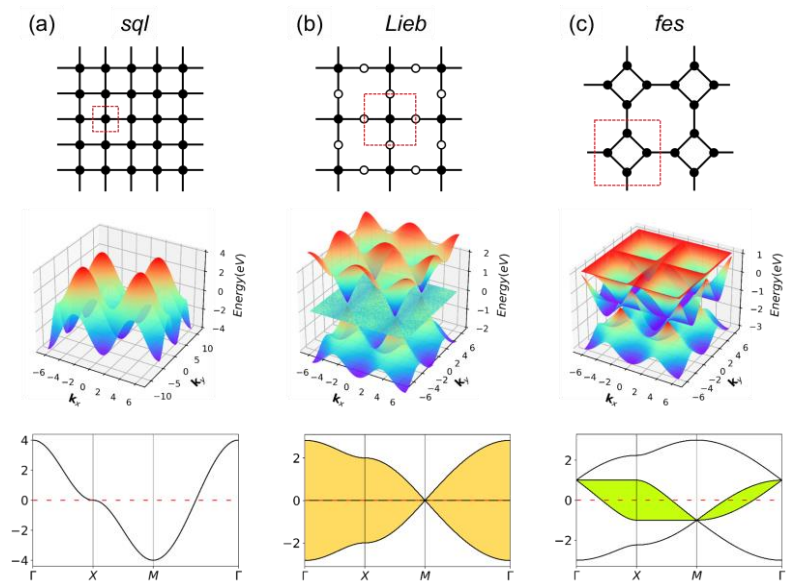

Figure 1. Schematic models and band structures of (a) *sql*, (b) *Lieb*, and (c) *fes* lattices in the Tight-Biding model, considering only 1st-neighbor interactions. The red dashed lines indicate the unit cells. Yellow areas indicate *Lieb*, green areas *fes* bands.

So far, *Lieb* lattice has been investigated mainly by using optical lattices,[17] or by surface deposition of small molecules.[18] A recently synthesized two-dimensional (2D) $sp^2$ carbon-conjugated covalent organic framework (COF)[19] was theoretically demonstrated to have *Lieb bands*.[13,16] Soon after, the ZnPc polymer, an analogue to the experimentally achievable FePc polymer,[20] with zinc-phthalocyanine as the lattice center and benzene ring as the linker, was predicted to have *Lieb bands*, which remain topologically non-trivial after chemical substitution or physical strain engineering.[15] However, while ZnPc band structure was identified as a *Lieb* lattice in the original paper, it much more resembles that of *fes lattice*, which has been in depth studied in refs [11,21]. *Fes* characteristic bands (denoted as *fes bands* in the remaining manuscript, Figure 1c), have two high-symmetry crossing points ($\Gamma$ and $M$), with one locally flat band and two conical bands crossing at a Dirac point.[10,11]

In this study, we have investigated a series of hypothetical two-dimensional polymer (2DP) structures derived from the zinc-phthalocyanine (ZnPc) 2DP as model *sql* polymers. These derivatives, denoted as ZnPc-*x*P, feature linkers with varying lengths, where *x* represents the number of aromatic rings in the linker (ranging from 1 to 5, Figure 2).[22] Interestingly, the band structure of the ZnPc-*x*P 2DPs, while structurally having a simple *sql* lattice, exhibits an evolution from *fes bands* to *Lieb bands*, depending on the linker chain length. In particular, the *Lieb bands* of ZnPc-4P material are in perfect agreement with the TB model of *Lieb* lattice, including their non-trivial electronic topology. We have also shown that the features of the *Lieb bands* are preserved when the Fermi level is shifted by both simple electron removal or atom substitution, thereby transforming the challenge of *Lieb* lattice search into a well-defined chemical synthesis task.

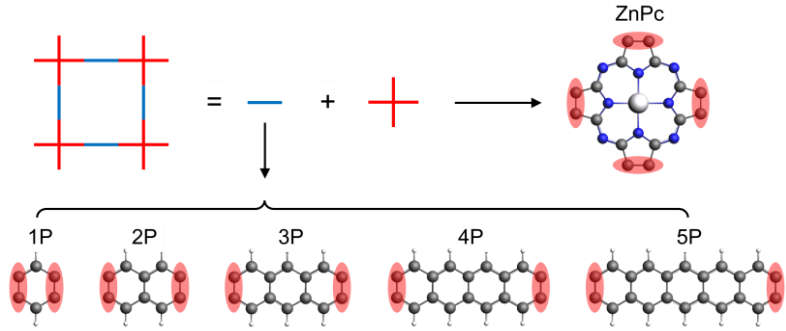

Figure 2. Schematic representation of ZnPc-*x*P 2DPs series structure. The red shaded area indicates the shared atoms between the center and linker molecules.

**RESULTS**

The basic building units of the 2DPs of interest in this study are the ZnPc molecule, and acenes (benzene, naphthalene, anthracene, tetracene, and pentacene, denoted as 1P, 2P, 3P, 4P, and 5P, respectively). These building blocks assemble into flat 2D sheets with square pores, forming ZnPc-*x*P 2DPs, where $x=1, 2, \ldots, 5$ represents the number of aromatic rings in the linker. The ZnPc-*x*P 2DPs family represents a very interesting case study into the lattice structure and electronic properties relationship. From the structural point of view, common in experimental/synthetic materials community, they would be considered as *sql*. However, in a deeper look, *Lieb* and *fes* lattices can also be formally projected on the ZnPc-*x*P geometries (Figure 3). Thus, we have investigated the relationship between ZnPc-*x*P 2DP structures and electronic properties, particularly considering them as possible materials possessing *Lieb bands*.

Indeed, the band structures of ZnPc-*x*P 2DPs shown in Figure 4 include both *fes bands* and *Lieb bands* features. The *fes bands* are prominently present in the band structure of ZnPc-1P 2DP. The *Lieb bands* can also be observed below the Fermi level (set to 0) with a distorted flat band located at about -2.3 eV. With increasing linker length, the (almost) flat band gradually approaches the Fermi level, as indicated by the yellow arrow in Figure 4. The *Lieb bands* are slightly distorted because the flat band has a small dispersion. It is noteworthy that the position of the Dirac cone alternates between $\Gamma$ and $M$ points as the linker length changes in ZnPc-*x*P 2DPs (Figure 4 which originates from the parity of the symmetry. The parity in 2DPs structures has recently been reported in the same material by Raptakis et al.[22] Upon further investigation, we found that the same behavior can be observed in other classes of square polymers, as shown in the model structures in Figure SI-1.

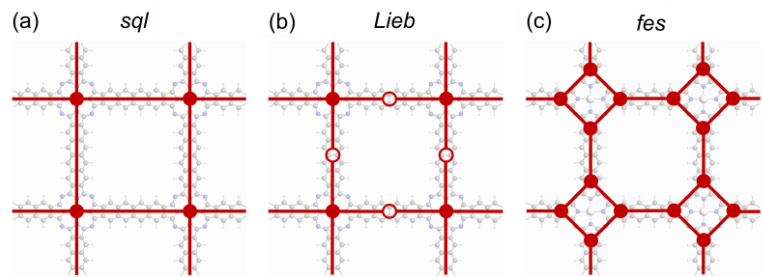

Figure 3. Possible topologies (a) sql, (b) Lieb, and (c) fes lattices projectable on the ZnPc-xP 2DPs. ZnPc-4P is shown as an example.

The contribution of the center/linker to the band structure of ZnPc-*x*P 2DPs is shown in Figure 4 (*center* defined as a porphyrin molecule equivalent and *linker* as the connectors in between). The flat band of *Lieb bands* is mainly contributed by the linker, and the Dirac cone bands are contributed by both the center and the linker. The *fes bands* in ZnPc-1P are dominated by the center with only a small contribution from the linker. The contribution varies depending on the definition of the cen-ter/linker region in the 2DP structures, as shown in Figure SI-2, but shows the same trend. As the linker length increases, the contribution of the linker to the bands forming the Dirac cone gradually increases, while the *fes* lattice features become less pronounced.

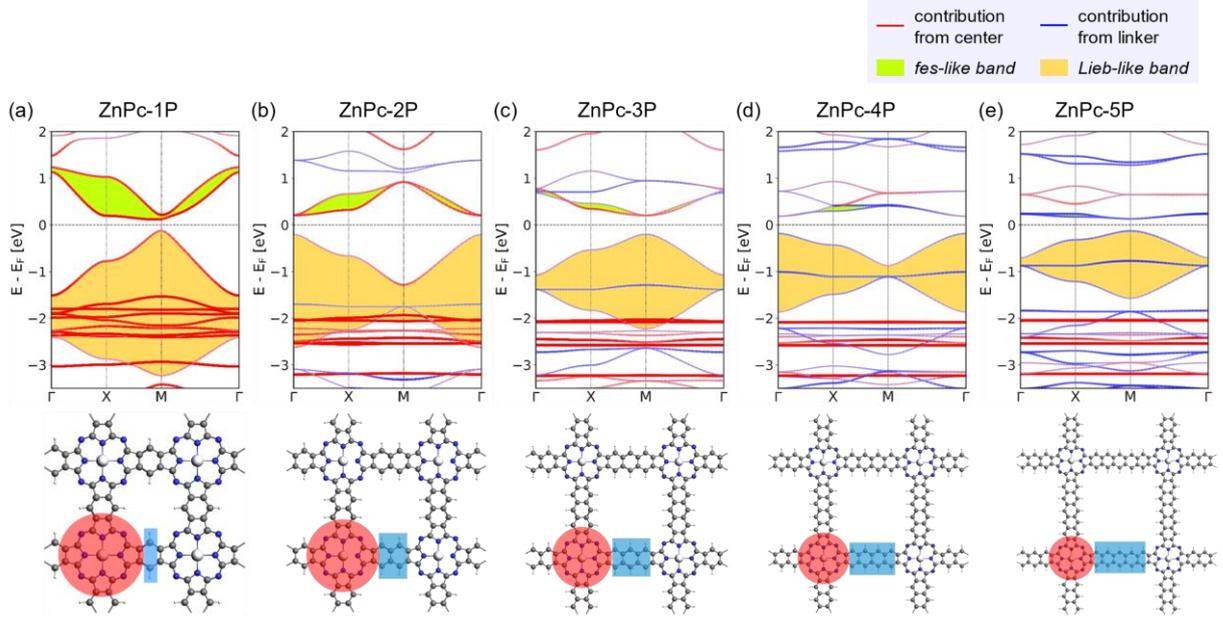

Figure 4. Band structure and the top view of the ZnPc-$x$P 2DPs of (a) ZnPc-1P, (b) ZnPc-2P, (c) ZnPc-3P, (d) ZnPc-4P, and (e) ZnPc-5P. The red and blue bands indicate the orbital contribution from the center/linker, respectively, which are also highlighted in the crystal structure using the same color scheme. *fes bands* and *Lieb bands* features are highlighted with green and yellow backgrounds in the band structures.

The contribution of the center/linker to the band structure of ZnPc-$x$P 2DPs is shown in Figure 4 (*center* defined as a porphyrin molecule equivalent and *linker* as the connectors in between). The flat band of *Lieb bands* is mainly contributed by the linker, and the Dirac cone bands are contributed by both the center and the linker. The *fes bands* in ZnPc-1P are dominated by the center with only a small contribution from the linker. The contribution varies, depending on the definition of the center/linker region in the 2DP structures, as shown in Figure SI-2, but shows the same trend: as the linker length increases, the contribution of the linker to the Dirac cone bands gradually increases, while the *fes* lattice features become less pronounced.

To further verify that ZnPc-$x$P 2DPs can be regarded electronically as a *Lieb* lattice, we have fitted the band structure to a TB model of a *Lieb* lattice to reproduce the electronic structure of ZnPc-4P. The TB model has four parameters: on-site energies for the edge and center sites, and hopping parameters amongst the nearest neighbors. The parameters of the model were optimized to fit the *Lieb bands* of ZnPc-4P. Our analysis shows an excellent match between the band structures from the TB model and the bands from full ZnPc-4P (Figure 5).

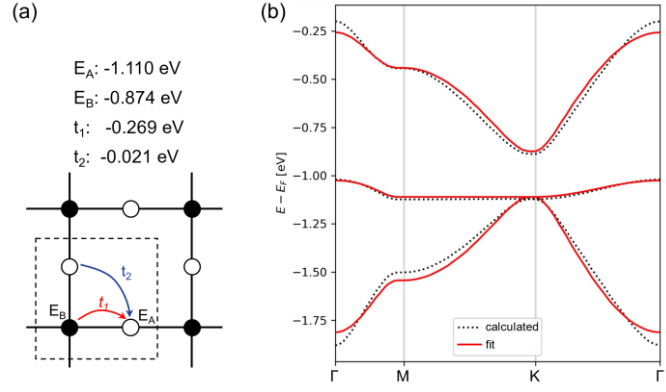

Figure 5. (a) *Lieb* lattice with four parameters, the center site $E_A$, the corner site $E_B$, the nearest neighbor hopping $t_1$ and the next-nearest neighbor hopping $t_2$. (b) the TB fitting of DFT calculated band structures.

In order to understand how the electronic topology originates from the atomistic structure, we have constructed three simplistic, full-atomic hypothetical carbon allotrope models featuring *fes* and *sql/Lieb* lattices similar to the TB lattice models. The first model (Figure 6a), T-graphene,[23] constitutes exactly *fes* lattice, and produces nearly ideal fes bands. The electronic features of the *fes* lattice are preserved in T-graphtriyne[24] (Figure 6b), which incorporates an extended linker between the rhombic nodes. If the rhombic center of T-graphtriyne is substituted with a single Zn atom forming Zn-diyne (Figure 6c), which would traditionally be considered a *sql* lattice, distorted *Lieb bands* emerge instead of the *sql* bands.[10] These simple models confirm that *sql* TB model is too simplistic to describe square-pore 2DP systems, since even as simple $D_{4h}$ symmetry structure as in Figure 6c contains *Lieb bands*. The reason for this is that the electronic structure is defined by topology of the scalar field of electron density rather than by simple geometry, which means that the lattice of the material cannot be simply mapped by its atomistic structure.

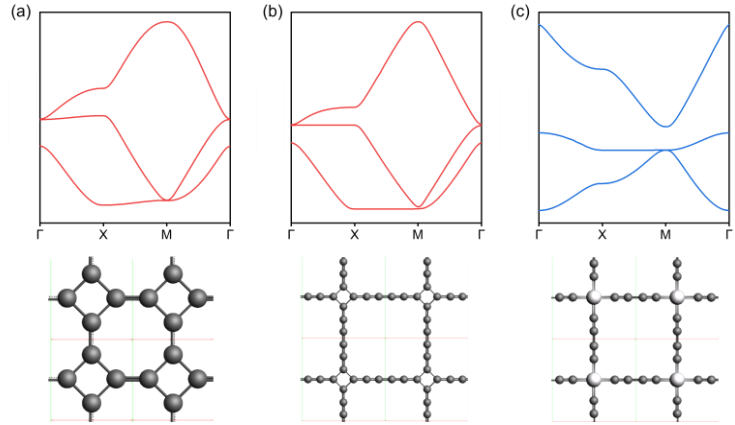

Figure 6. Band structures and the top view of model structures connected by (a) T-graphene (b) T-graphtriyne, and (c) Zn-diyne. Features of *fes* and *Lieb* in the band are colored in red and blue, respectively.

We have evaluated the topological properties of the ZnPc-*x*P materials by calculating their Chern numbers, taking into account spin-orbit coupling (SOC). The Chern numbers of the three *Lieb bands* are (-1 1 0) from the bottom band to the top band, which means that the flat band and the bottom Dirac band show non-trivial topological properties. The fitted TB model for ZnPc-4P shows the same Chern numbers. The topological properties often depend on the strength of the SOC effect structure, which can be modulated e.g. by incorporating heavy metal atoms into the structure. We have investigated this on the fitted TB model of ZnPc-4P by scanning the effect of artificially set SOC values. Under heavy SOC effects, robust edge states can emerge in the electronic structure (Figure SI-3, 4).

The primary limitation of the studied 2DPs lies in the positioning of the *Lieb bands* below the Fermi level. Achieving non-trivial structural properties in these bands thus requires shifting the Fermi level while preserving the *Lieb bands*. To address this, we first conducted an analysis of the atom contributions to the charge density within the *Lieb bands* (VB1, VB2, and VB3) using ZnPc-4P 2DP as a model system (Figure 7a, Figure SI-5). The primary contribution to the *Lieb bands* comes from the linker (tetracene) π-electrons, which contribute to all VB1-3, albeit the carbons in the phthalocyanine center also partially contribute to the Dirac bands VB1 and VB3 (Figure 7).

The contribution of the metal atom to the *Lieb bands* is only minimal. This is also confirmed using a structure without a metal atom in phthalocyanine center (Pc-4P), which gives almost identical *Lieb bands* as ZnPc-4P (Figure 7). However, the *fes* features in conduction bands collapse after the removal of the metal atom.

Additionally, we have designed a hypothetical reference porphyrin polymer characterized by an "ideal atomic Lieb lattice" structure ZnPc-ZnPc 2DP (Figure 7c, Figure SI-6). Both fes features in conduction bands and Lieb bands in valence bands are present with only minor changes. Lieb bands again contain the full π-system of the linker, despite its bigger size. This suggests that the well-ordered conjugated π-system is important for achieving high quality Lieb electronic materials.

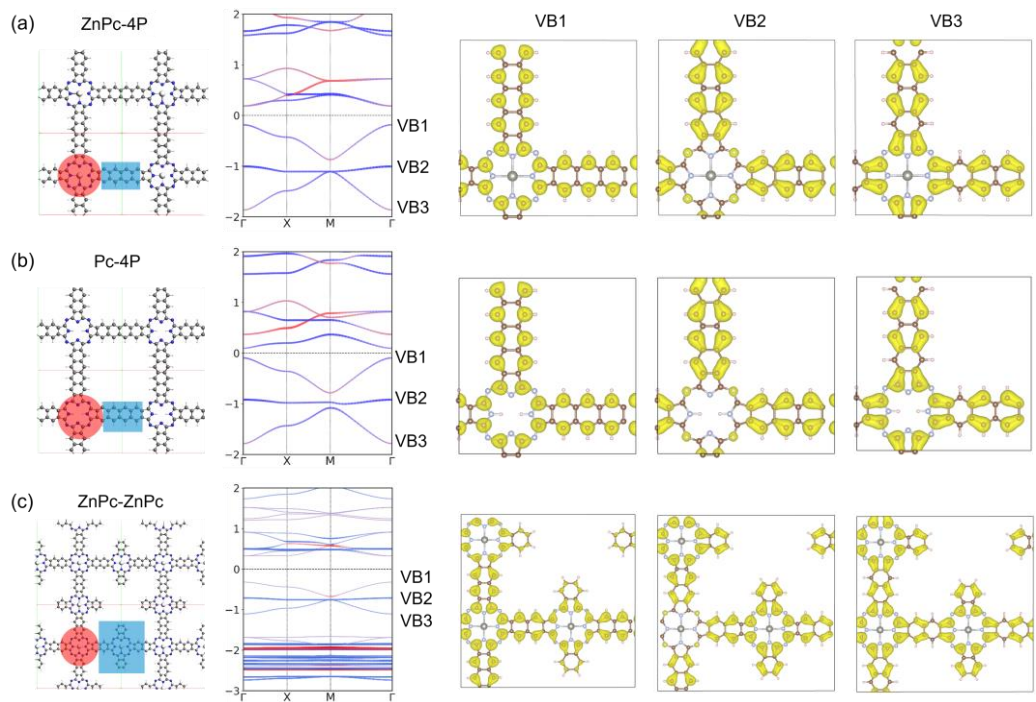

Figure 7. Structure, band structure, and charge density distribution at $\Gamma$ point of the upper three VB bands of (a) ZnPc-4P with Zn atom, (b) Pc-4P without Zn atom, (c) ZnPc-ZnPc 2DP. The red and blue bands indicate the orbital contribution from the center/linker, with the center-linker partitioning highlighted in the crystal structure using the same color scheme.

We have investigated the modulation of the Fermi level position by two different methods: the direct removal of electrons from the system and atomic substation.[25] The removal of two electrons per unit cell effectively shifts the Fermi level toward the flat band. This adjustment results in a slight increase in the dispersion of the flat band, but leaves the fundamental *Lieb bands* intact (Figure 8). The charge density contributions from the *Lieb bands* show identical features to those observed in the pristine structure. The Chern number of CB1 is 2, showing non-trivial topological properties, while it is 0 for VB1 and VB2.

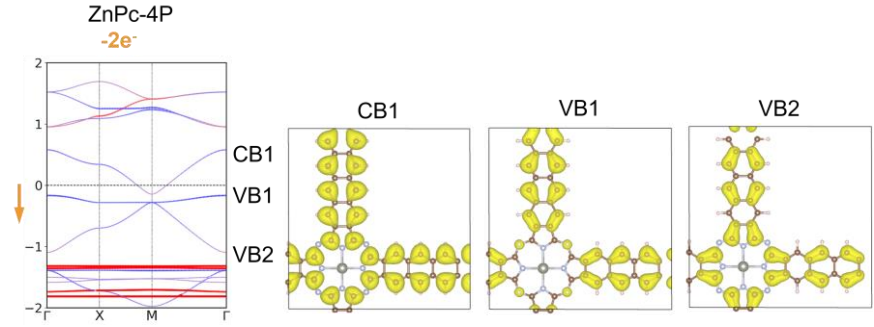

Figure 8. Band structure and charge density distribution at $\Gamma$ point of the three *Lieb bands* of ZnPc-4P with two electrons removed (two holes introduced) per unit-cell. The red and blue bands indicate the orbital contribution from the center/linker.

The charge density analysis of VB1-3 (Figure 7) shows that in order to access the flat band, electrons should be removed from the aromatic system of the linkers, especially in the edge atoms of the linker. We have tested replacing individual carbons (2 atoms per UC, one per linker) in the linker with boron atoms, shifting the Fermi level to the flat band; and with nitrogen, shifting the Fermi level to the *fes* bands (Figure 9). In the B-substituted structure, the Fermi level shifts to the *Lieb bands*, and most of the band structure features are preserved, although the dispersion of the "flat band" is stronger. The two Dirac bands show non-trivial topological properties with Chern number of (-1 0 1) for the *Lieb bands*. In the N-substituted structure, the *Lieb bands* are preserved, with an additional flat band crossing. These substituted model structures, although not easily achieved chemically, could provide a general guide how *Lieb bands* could be accessed in other square pore 2DPs with more suitable structures for substitution.

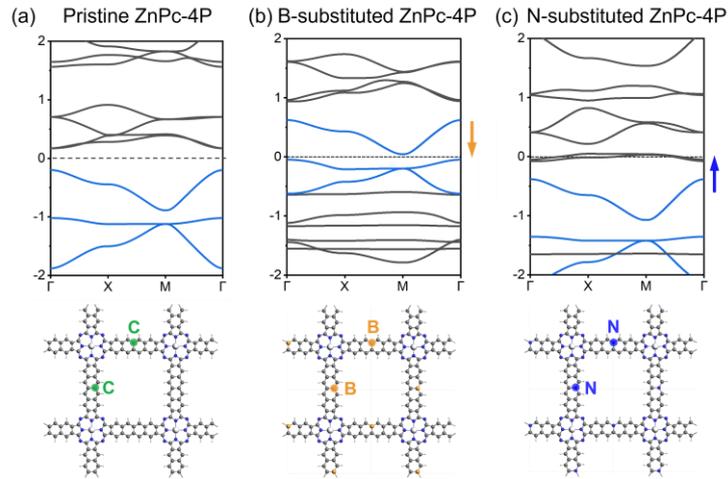

Figure 9. Band structures, and top view of the structures of (a) pristine (b) B-substituted, (c) N-substituted ZnPc-4P. Lieb bands are colored in blue. The arrow indicates the Fermi level shift direction.

**Conclusions**

We have investigated the ZnPc-*x*P 2DPs as a model system for electronic structure topology of square pore 2DPs. While these materials are traditionally considered to have a *sql* topology well known from TB models, our results challenge this over-simplified view. We found that ZnPc-*x*P 2DPs exhibit both *fes bands* and *Lieb bands* while completely lacking the expected *sql* features. This is because the electronic structure of these materials is governed by the topology of the electron density, rather than by simple atomic geometry. Furthermore, the flat band and bottom Dirac band of the *Lieb bands* possess non-trivial topological properties, as evidenced by non-zero Chern numbers. Unfortunately, the *Lieb bands* in the ZnPc-*x*P 2DPs are lo-cated below the Fermi level, as well as in most other model materials studied here. However, we have shown that shifting the Fermi level by controlling the number of electrons in the system via gating or substitution preserves the *Lieb bands* including their topological character. The challenge of finding *Lieb* lattice structures thus turns into a well-defined chemical synthesis task. This opens up possibility of designing materials with unique properties, such as topologically non-trivial phase. We hope that our work will stimulate further experimental exploration of *Lieb*-lattice-based topological materials.

### Methods

The geometries were optimized using the self-consistent-charge density functional based tight binding (SCC-DFTB)[26] method as implemented in the Amsterdam Modelling Suite (AMS) ADF 2019.[27] The 3ob-3-1 parameter set[28] was used for systems with X-Y element pair interaction (X, Y = C, H, Zn), and the matsci-0-3 parameter set was applied for systems including boron.[29] Band structure calculations were performed employing FIH-aims with TIER1 basis set with 4×4×1 k-mesh grid[30], using DFT with the Perdew, Burke, and Ernzerhof (PBE) functional.[31] The key parameter "tight" regarding computational accuracy was used to control all integration grids, and the accuracy of the Hartree potential. The Berry curvature and the intrinsic anomalous Hall conductivity were performed using the WANNIER90 package.[32] Calculations of edge states and the Chern number were carried out using WannierTools package.[33] SOC and spin polarization were taken into account in the topological calculations. The geometries and corresponding band structures for all systems investigated in this work are available at the NOMAD repository as a dataset under doi: 10.17172/NOMAD/2023.11.26-1.

### Acknowledgements

Y. Z. acknowledges Tsai-Jung Liu, Florian Arnold and Dr. Hongde Yu for fruitful discussions. Y.Z. acknowledges China Scholarship Council. S.Z. acknowledges funding of DFG priority program SPP 2244. M. P. is supported by project EMPIR 20FUN03 COMET; this project has received funding from the EMPIR programme co-financed by the Participating States. All authors thank Deutsche Forschungsgemeinschaft for support within CRC 1415. All authors thank computational resources provided by ZIH Dresden and the NHR Center PC2.

# Electronic Lieb lattice signatures embedded in two-dimensional polymers with square lattice

Yingying Zhang,[a] Shuangjie Zhao,[a] Miroslav Položij,[abc] Thomas Heine*[abcd]

[a.] Chair of Theoretical Chemistry, Technische Universität Dresden, Bergstrasse 66, 01069 Dresden, Germany

[b.] Helmholtz-Zentrum Dresden-Rossendorf, Bautzner Landstr. 400, 01328 Dresden, Germany

[c.] Center for Advanced Systems Understanding, CASUS, Untermarkt 20, 02826 Görlitz, Germany

[d.] Department of Chemistry and ibs for nanomedicine, Yonsei University, Seodaemun-gu, Seoul 120-749, Republic of Korean.

E-mail: thomas.heine@tu-dresden.de

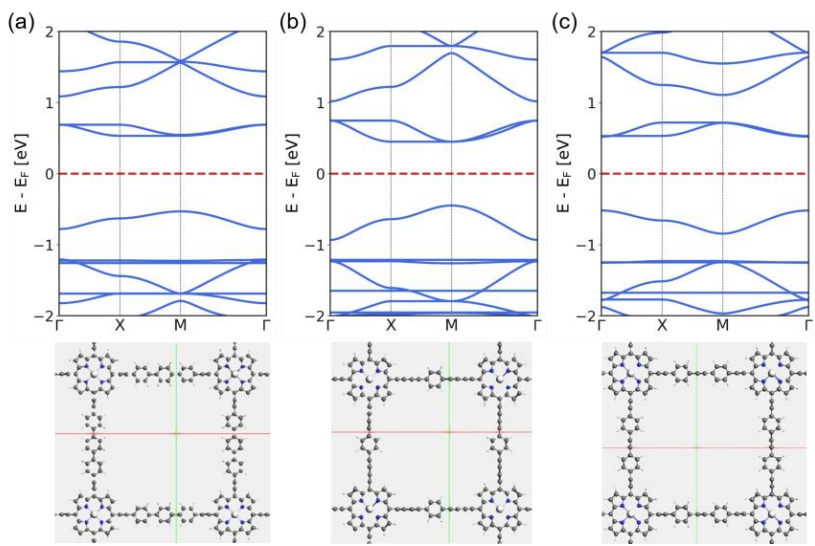

Figure SI-1. Band structure and the top view of the ZnPc-xP COF, connected with porphyrin and phenyl group. (a) ZnPP-3P COF, (b) ZnPP-1P COF, and (c) ZnPP-2P COF, showing the parity of the Dirac cone position, altered between M and Γ.

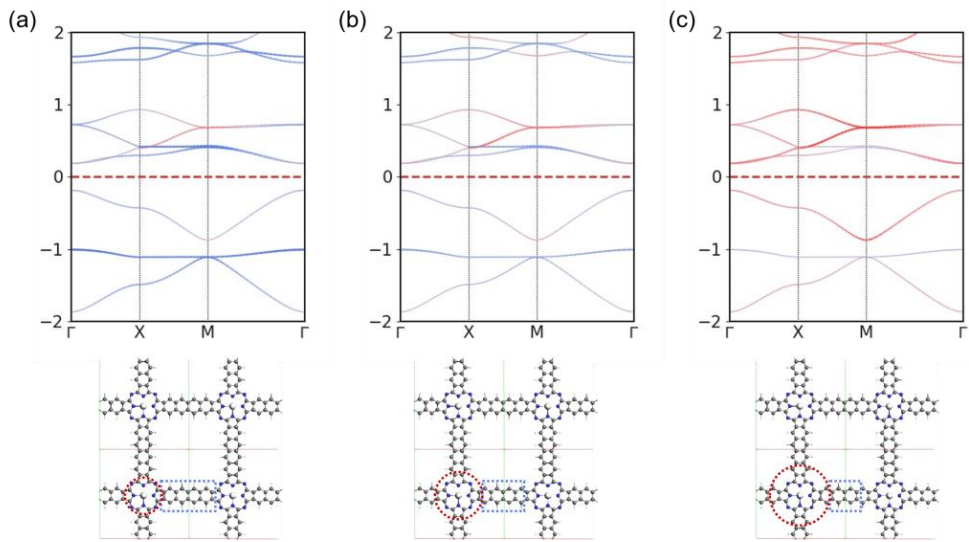

Figure SI-2. Band structure and the top view of the ZnPc-4P COF in different boarder position between center/linker. The red and blue bands indicate the orbital contribution from center and linker, which are highlighted with the same color in the crystal structure.

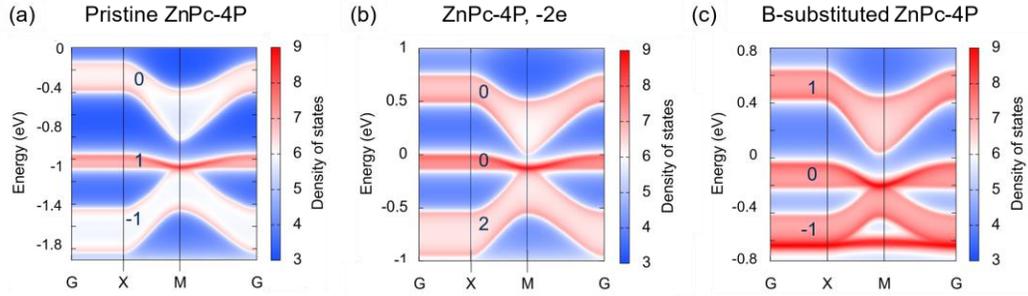

Figure SI-3. Edge states and Chern number of (a) pristine ZnPc-4P, (b) ZnPc-4P with two electrons removed (two holes introduced) per unit-cell, (c) B-substituted ZnPc-4P.

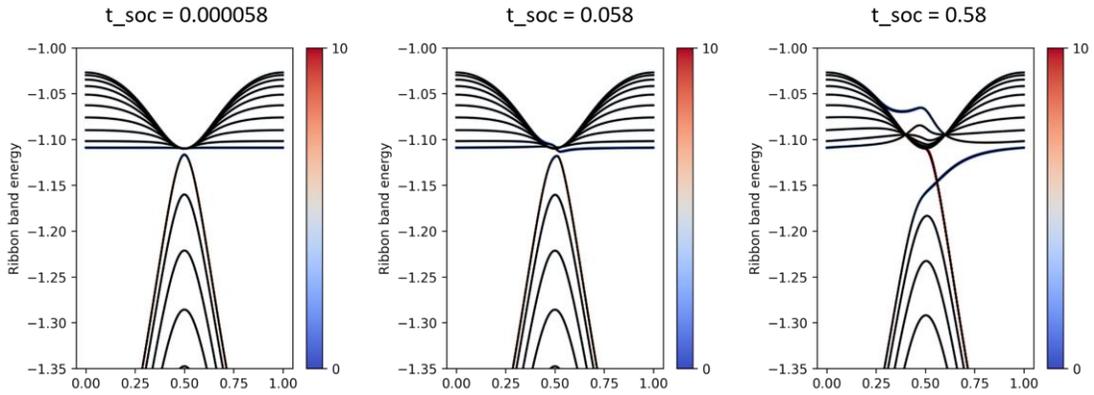

Figure SI-4. Edge states of the fitted TB model of ZnPc-4P at different degrees of spin-orbit coupling (SOC).

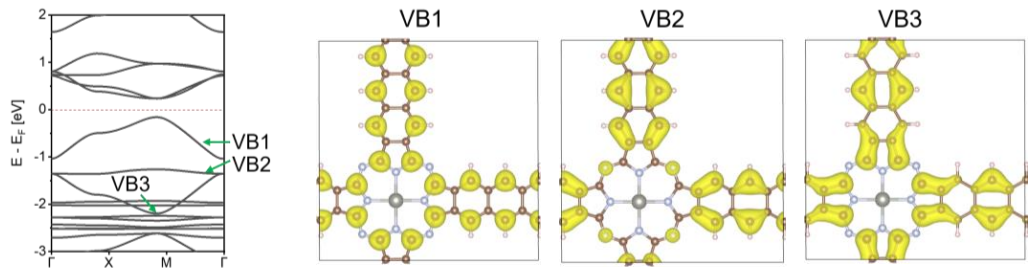

Figure SI-5. Band structure, and charge density distribution at $\Gamma$ point of the top three VB bands of ZnPc-5Bz.

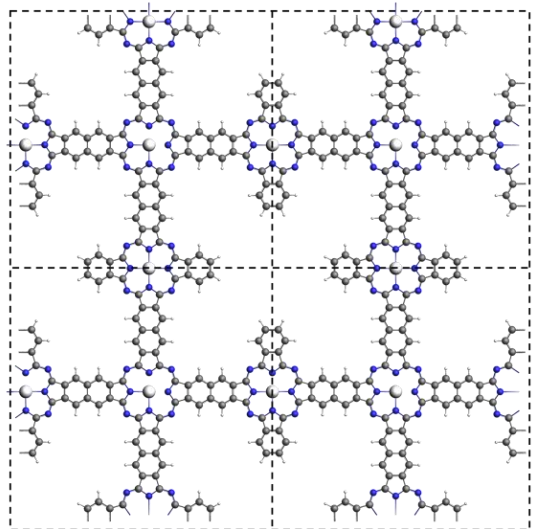

Figure SI-6. Top view of ZnPc-ZnPc COF.